\documentclass[a4paper,12pt]{article}
\usepackage{amssymb,amsfonts,amsmath}

\usepackage{graphicx} %
\graphicspath{ {images(bw)/} }%

\begin{document}
\begin{center}
{\large\bf Chiral imbalance in hadron matter : its manifestation in photon polarization asymmetries}

{A.A.Andrianov$^{1,2,\P}$},  {V.A.Andrianov$^{1}$}, {D.Espriu$^{2}$}, {A.V.Iakubovich$^{1}$}, {A.E.Putilova$^{1}$}

$^1${Faculty of Physics, Saint Petersburg State
University,  Saint Petersburg 199034, Russia}

$^2${Departament de F\'isica Qu\`antica i Astrof\'isica and Institut de Ci\'encies del Cosmos (ICCUB),Universitat de Barcelona, 08028 Barcelona, Spain}

$^\P${E-mail: a.andrianov@spbu.ru}
\end{center}

\centerline{\bf Abstract}
The possibility of the formation of a local parity breaking (LPB) in a quark-hadron medium is examined as a result of  chiral symmetry violation, i.e, the appearance of difference between the average densities of right- and
 left-handed quarks in the fireball after heavy-ion collisions (HIC). Using the effective meson Lagrangian motivated by QCD extended to the chiral medium the properties of light scalar and pseudoscalar mesons are analyzed. It is establish that exotic decays of scalar mesons arise due to mixing of $\pi$ and $a_0$ vacuum states in the presence of chiral imbalance. The pion electromagnetic formfactor obtains a parity-odd supplement which generates a photon polarization asymmetry in pion polarizability. We conjecture that the above-mentioned properties of LPB can be revealed in experiments on LHC, RHIC, FAIR and NICA accelerators.

\section{ Introduction: Chiral Imbalance}
This work concerns the problem of CP- violation in the strong interactions in HIC at high energies: LPB in the fireball with a chiral quark  matter namely,  in the presence of Chiral Imbalance, the difference between the average densities of right and left- handed quarks. After hadronization of quark-gluon medium, the fireball mostly consists of the light meson gas. In such a medium mesons are composed from nearly massless $ u $ and $d $ quarks and they change their properties, as compared to vacuum ones. In particular, in heavy-ion collisions at high energies, with raising temperatures and baryon densities, a collective metastable state can appear in the fireball with a non-trivial topological/axial charge, which is related to fluctuations of gluon fields. As described earlier \cite{aaep} in such an environment  the emergence of a gradient density of isosinglet pseudoscalar condensate is possible which can be formed as a result of large, “long-lived” topological fluctuations of gluon fields in a fireball in central collisions  \cite{aaep}. To describe  such a quark-hadron matter with chiral imbalance in a fireball one can introduce the axial/chiral chemical potential \cite{aaep}. There are some experimental indications of an abnormal dilepton excess in the range of low invariant masses and rapidities, and of moderate values of transverse momenta \cite{00}, which can be partially thought of as a result of LPB in the medium (the details can be found in \cite{tmf}).

We are going to describe new properties of the light mesons in a chiral imbalance medium.
The work includes the following items:
finding of modified vertices of meson decays;
obtaining of the decay widths dependence on energy and chiral chemical potential;
consideration of a pion-photon scattering in chiral medium;
definition and calculation an expression for the polarization asymmetry in a pion-photon scattering.

The behaviour of baryonic matter under extreme conditions has got
recently a lot of interest \cite{1}. A medium generated in the heavy
ion collisions may serve for detailed  studies, both experimental
and theoretical, of various phases of hadron matter. In this context
new properties of QCD in the hot and dense environment are tested in
current accelerator experiments on RHIC and LHC \cite{2}.

In heavy ion collisions, in principle, there are two distinct
experimental situations for peripheral and central collisions. In
the first case  the so-called Chiral Magnetic Effect(CME) can be
detected, details see in ~\cite{kpt} and also ~\cite{6} for a review
and additional references.

In the second case there are some experimental indications of an
abnormal dilepton excess in the range of low invariant masses and
rapidities and moderate values of the transverse momenta ~\cite{7}
(see the reviews in ~\cite{tser}), which can be thought of as a
result of LPB in the medium (the details can be found in
~\cite{tmf}). In particular, in heavy-ion collisions at high
energies, with raising temperatures and baryon densities, metastable
states can appear in the finite-volume fireball with a nontrivial finite-volume
"topological" charge (due to fluctuations of gluonic fields)
~$T_{5}$.

Based on the PCAC relation one can find the ratio of a nonzero topological
charge with a non-trivial quark axial charge ~$Q^q_{5}$ ~\cite{tmf}.
Namely, integrating
over a finite volume of fireball we come to the relation,
\begin{equation}
\label{eq23}
\frac{d}{dt}(Q_5^q-2 N_f T_5) \simeq 0,
\quad
Q_5^q=\int_{\mbox{\rm vol.}}d^3x\,q^\dagger\gamma_5 q = \langle
N_{L}-N_{R}\rangle,
\end{equation}
where $\langle N_{L}-N_{R}\rangle$ stands for the vacuum averaged
difference of left and right chiral densities of baryon number ({\it Chiral Imbalance}).
Therefrom  it follows that in the chiral limit the
axial quark charge is conserved in the presence of non-zero
(metastable) topological charge. If for the lifetime
and the size of hadron fireball of order $L=5-10$~fm , the average
topological charge is non-zero, $\langle \Delta T_5 \rangle\ne 0$,
then it may be associated with a topological chemical potential
$\mu_T$
or consequently with an axial chemical potential~$\mu_5$ ~\cite{aaep} for  light~$u$,~$d$ quarks. Thus we have,
\begin{equation}
\langle \Delta T_5 \rangle \simeq \frac{1}{4} \langle
Q_5^q\rangle\, \Longleftrightarrow\,\mu_5 \simeq
\frac{1}{4}\mu_T, \label{eq24}
\end{equation}
Hence adding to the QCD lagrangian the term $\Delta{\mathcal L}_q=\mu_5
Q_5^q$, we simulate in QCD the possibility of accounting for non-trivial
fluctuations of topological charge (fluctons) in the nuclear (quark) fireball.

Thus in a quasi-equilibrium situation the appearance of a nearly
conserved chiral charge can be incorporated with the help of an
axial (chiral) vector  chemical potential $b_\mu$. The appearance of
a space vector part in $b_\mu$ can be associated with the
non-equilibrium axial charge flow \cite{kharzeev}.

\section{QCD-inspired effective meson Lagrangian for $ SU_f (2)$ case}

One can reckon on the quark-hadron continuity ~\cite{fuku2018} and for the detection of Local Parity Breaking in the hadron fireball
implement the generalized sigma model with a background 4-vector of axial chemical potential ~\cite{sigma}, symmetric under
$ SU_{L}(N_{f})\times SU_{R}(N_{f})$, for $ u,d$-quarks ($N_{f}=2$),
\begin{align}
{L\over N_c}&=\cfrac{1}{4}\,\text{Tr}\,(D_{\mu}H\,(D^{\mu}H)^{\dagger})
+\cfrac{B}{2}\,\text{Tr}\,[ \,m(H\,+\,H^{\dagger})]
+\cfrac{M^{2}}{2}\,\text{Tr}\,(HH^{\dagger})
\nonumber \\
\,&-\cfrac{\lambda_{1}}{2}\,\text{Tr}\,\left[(HH^{\dagger})^{2}\right]
-\cfrac{\lambda_{2}}{4}\,[\,\text{Tr}\,(HH^{\dagger})]^{2}
+\cfrac{c}{2}\,(\det H+\,\det H^{\dagger}),
\label{lagr_sigma}
\end{align}
where $H = \xi\,\Sigma\,\xi$ is an operator for meson fields, $N_c$ is a number of colours,
\(m\) is an average mass of current $u,d$ quarks,
\(M\) is a "tachyonic" mass generating the spontaneous breaking of chiral symmetry,
\(B, c,\lambda_{1},\lambda_{2}\) are real constants.

The matrix \(\Sigma\) includes the singlet scalar meson \(\sigma\), its vacuum average \(v\) and the isotriplet of scalar mesons \( a^{0}_{0},a^{-}_{0},a^{+}_{0}\), the details see in ~\cite{sigma}.

The operator \(\mathbf \xi\) realizes a nonlinear representation of the chiral group and is determined by the isotriplet \( \pi^{0},\pi^{-},\pi^{+}\)
of pseudoscalar mesons,~\cite{sigma}.

The covariant derivative of $H$ contains external gauge fields
${R}_{\mu}$ and ${L}_{\mu}$,

$ D_{\mu}H= \partial_{\mu}H-i{L}_{\mu}H+iH{R}_{\mu}$.

These fields include the photon field \(A_{\mu}\) and are supplemented also by a background 4-vector of axial chemical potential \( (b_\mu) = (b_0, {\mathbf b})\),
$ {R}_{\mu} = e\,Q_{em}A_{\mu}- b_\mu \cdot 1_{2\times2} $,
$ {L}_{\mu} = e\,Q_{em}A_{\mu}+b_\mu \cdot 1_{2\times2} $,
where
\(Q_{em} = \frac{1}{2}\tau_{3}+\frac{1}{6}1_{2\times2}\)
is a matrix of electromagnetic charge.

The complete effective meson lagrangian has to include also a P-odd part: the Wess-Zumino-Witten effective action \cite{WZW} which is modified in the chirally imbalanced medium. The relevant parts of WZW action read,
\begin{equation}
\Delta{\cal  L}_{WZW}= -\frac{ie\,N_{c} b_\nu }{6\pi^{2}\,v^{2}}\,\epsilon^{\;\nu\sigma\lambda\rho}\,
A_{\rho}(\partial_{\sigma}\pi^{+})\,(\partial_{\lambda}\pi^{-})
-\frac{e^{2}N_{c}}{24\,\pi^{2}v}\,\epsilon^{\,\nu\sigma\lambda\rho}\,
(\partial_{\sigma}A_{\lambda})(\partial_{\nu}A_{\rho})\pi^{0}
\end{equation}

\section{Chiral condensate dependence on chiral chemical vector}
The mass gap equation for the scalar condensate follows from (\ref{lagr_sigma}),
\begin{equation}
-4\left(\lambda_{1}+\lambda_{2}\right)v^3
+\left(2M^2+4 b^2+2c\right)v+2B\,m=0 .\nonumber
\label{massgap}
\end{equation}
The general solution of this equation $v(b_\mu)$ can be found
explicitly ~\cite{epjqfthep17}.

Note that there are different regions for chiral vector, invariant
under Lorentz transformations of the fireball frame:
chiral imbalance region where $b^2 > 0$, then in the rest frame the chiral background $(b^\mu) = (\mu_5, 0,0,0)$;\\
chiral vector imbalance region where $b^2 < 0$, then in the static frame the chiral background can be along the beam axis $(b^\mu) = (0,0,0,b)$.

 In a hot and dense medium Lorentz invariance is broken by thermal bath and baryon density, thus the physical effects depend on a particular set of components of $(b^\mu)$.

 In the chiral imbalance region with $b^2 > 0$ the increasing of chemical potentials triggers the growth of chiral condensate, i.e. enhances Chiral Symmetry Breaking (CSB).\\
  Instead, in the chiral vector imbalance region with $b^2 < 0$ the chiral condensate is decreasing with growing $|b^2|$ up to
$|b^2| =\frac12 (M^2+c) $. At this scale in the chiral limit $m\to
0$ the CSB parameter $v \to 0$ and spontaneous CSB is
restored.

For $b^2 > 0$ in the rest frame of vector $(b_{\mu})= (\mu_5,0,0,0)$ one can compare  the predictions of our effective meson lagrangian
with the lattice estimations \cite{braguta}
that clearly shows the enhancement of CSB ~\cite{epjqfthep17}.

\section{Meson mass spectrum in chiral medium}

Introduce the definitions for meson state masses in the chiral imbalance environment.
The mass matrix for scalar and pseudoscalar mesons on the diagonal takes the following values, 
\begin{align}
m^{2}_{a}&=-2\left(M^2-2\left(3\lambda_{1}+
\lambda_{2}\right)v^2-c+2 b^{2}\right),
\nonumber\\
m^{2}_{\sigma}&=-2\left(M^2-6\left(\lambda_{1}
+\lambda_{2}\right)v^2+c+2b^{2}\right),
\nonumber\\
m^{2}_{\pi}&=\frac{2B\,m}{v}.
\end{align}
After diagonalization we define distorted masses as  $ m_{eff+}$ for
the field $ \tilde a$ and $ m_{eff-}$ for the field $ \tilde \pi $,
\begin{gather}
m^{2}_{eff\pm}=\frac{1}{2}\left( m^{2}_{a}+m^{2}_{\pi} \, \pm \, \sqrt{\left( m^{2}_{a}-m^{2}_{\pi} \,  \right)^{2}+(8\,b^\mu\,k_\mu)^{2}}  \right). \label{effmassgen}
\end{gather}

To fit the physical spectral data ~\cite{epjqfthep17}
 one can find
 $\lambda_{1}=1.64850\times10 $, $ \lambda_{2}=-1.31313\times10 $,
$ c=-4.46874\times10^4 $  $ \text{MeV}^2 $, $ B  =1.61594\times10^5 $ $\text{MeV}^2 $ .

After diagonalization of the mass matrix the states for $ a_0 $ and
$\pi$ mesons happen to be mixed. The eigenstates $ \tilde a, \tilde\pi$ are easily determined
~\cite {epjqfthep17}.

\section{P-odd vertices for light mesons}

Let us consider the  decay of the neutral pion into two photons, $\pi \to \gamma\gamma$. The expression for the differential decay width is given by relation (\ref{2}). Here $k$ is a momentum of photon and $q$ is a momentum of pion; $\theta$ is angle between $k$ and $q$; $f_{-}$ is a normalization of wave packets (cf. relation (\ref{norm})); $C_{+}$ is a diagonalization coefficient (cf. \cite{111,1111} ). The connection between $k$ and $q$ is given by expression (\ref{3}).

\begin{gather}
\frac{d\Gamma}{d\Omega}\negmedspace=\negthickspace
\left(
\!\frac{e^{2}N_{c}}
{24\pi ^{2}\,f_{\pi}}\!
\right)^{\negthickspace2}\negthickspace
\cdot
\frac{C^{\,2}_{+}(q_{0}\!)
|\mathbf{k}|^3
\negmedspace\sqrt{|\mathbf{k}|^2\!+\!|\mathbf{q}|^2
\!-\!2|\mathbf{k}||\mathbf{q}|\!\cos\theta}}
{\pi^{2}f_{-}(q_{0}\!)
\Bigg|
1\!
+\!\frac{|\mathbf{k}|-|\mathbf{q}|\cos\theta}
{\sqrt{|\mathbf{k}|^2+|\mathbf{q}|^2
-2|\mathbf{k}||\mathbf{q}|\cos\theta}}
\Bigg|}\:
\negthickspace\cdot\negthickspace
\left(
\!1\!
+\!\frac{|\mathbf{k}|
\!-\!|\mathbf{q}|\cos\theta}
{\sqrt{|\mathbf{k}|^2\!+\!|\mathbf{q}|^2
\!-\!2|\mathbf{k}||\mathbf{q}|\!\cos\theta}}\!
\right)^{\negthickspace2}
\label{2}
\end{gather}

\begin{equation}
|\mathbf{k}|=\frac{m^{\,2}_{e\!f\!f+}\!
\left(\!\sqrt{|\mathbf{q}|^2
\!+ \!m^2_{\text{eff+}}}+|\mathbf{q}|\cos\theta \right)}
{2\left(m^2_{\text{eff+}}\!+\!|\mathbf{q}|^{2}\!\sin^{2}\!\theta\right)}
\label{3}
\end{equation}

\begin{equation}
 f_{-}(q_{0}\!)\!=\!2q_{0}\!\!+
 \frac{32\,q_{0}\mu_{5}^{2}}
 {\sqrt{\!
 (m^{2}_{a}-m^{2}_{\pi})^{2}
 +\!(8\,q_{0}\mu_{5})^{2}}}
 \label{norm}
 \end{equation}

The $m_{\text{eff+}}$ is an effective mass of $a$ meson defined above.

After diagonalization of mass matrix for scalar and pseudoscalar mesons we define distorted masses as  $ m_{eff+}$ for the field $ \tilde a$ and $ m_{eff-}$ for the field $ \tilde \pi $ (cf. relation (\ref{effmassgen})).
The modulus of photon space momentum is obtained from energy conservation law. The width depends on three variables: space momentum modulus of photon, chiral chemical potential and the angle between directions of pion and photon momenta. The width can be integrated over the angle with some boundary values.
At zero chiral chemical potential the width value as it is known equals to 7.73 eV. Our formula gives 7.8 eV at $\mu_5$ equal to zero and the angle equal to $\pi$.

\begin{figure}[!h]
\center{\includegraphics[width=0.5\linewidth]{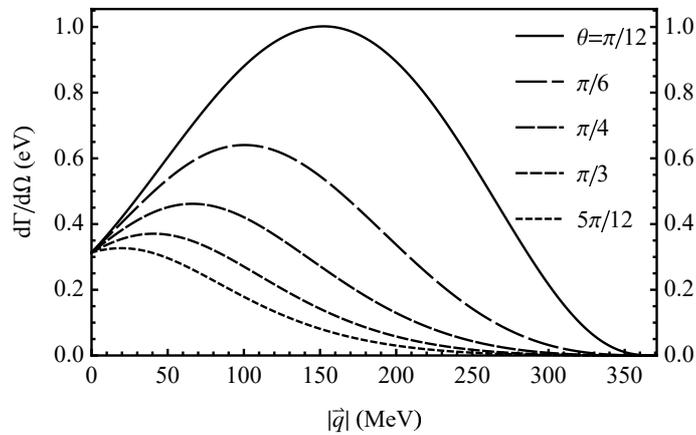} }
\caption{Differential decay width at \(\mu_{5}=100\) MeV}
\label{fig:1}
\end{figure}

The figure \ref{fig:1} shows the set of plots corresponding to different values of the angle. The chiral chemical potential is equal to 100 MeV. As we can see, the width has a maximum that unevenly decreases from 1 to 0.3 eV  with the angle increasing.

After mixing the decay ${\tilde a}^{\pm}_0\rightarrow \tilde\pi^{\pm}\gamma$ arises which breaks space parity.

\begin{figure}[!ht]
\centering
\includegraphics[scale=1.3]{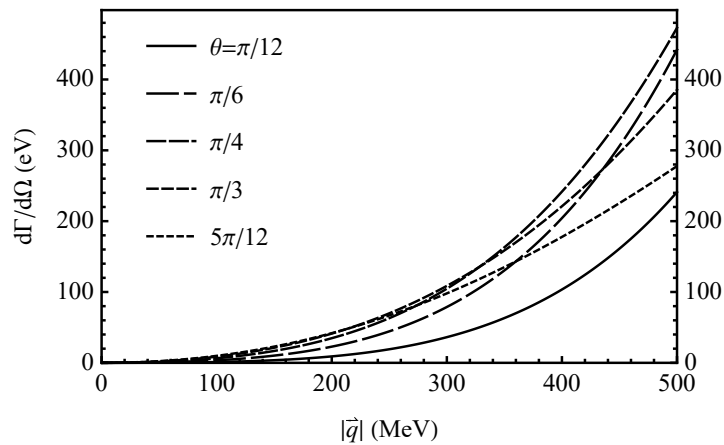}
\caption{Decay width $a^{\pm}\rightarrow\pi^{\pm}\gamma$, $\mu_{5}=100$ MeV}
\label{fig:width1_2d}
\end{figure}
\noindent
$q =|\vec q|$ is a space momenta of scalars. On this plot  the contribution of WZW vertex to the decay  is shown after the states mixing.
For pseudoscalars and scalars in flight  the speed of decays is considerably increasing.

\section{Pion-photon scattering and asymmetry in photon polarizations}
Let us consider the pion-photon scattering, where the intermediate state can be both a virtual pion and an $a_0$-meson.
Two parity-conjugated processes in (\ref{p_ph-p_ph}) differ in the polarization of the emitted photons.  Due to chiral symmetry violation, an asymmetry of the number of these processes can be observed. The order of asymmetry is such that the effect should be visible on modern accelerators of elementary particles. Therefore, this process could be one of the keys to finding strong CP-violation.
\begin{gather}
\pi^{\pm}( \vec p_{1} )
+\gamma (\vec k_{1} )
\rightarrow
\pi^{\pm}( \vec p_{2} )
+\gamma_{+}( \vec  k_{2} )
\nonumber \\
\pi^{\pm}( -\vec p_{1} )
+\gamma (-\vec k_{1} )
\rightarrow
\pi^{\pm}(-\vec p_{2})
+\gamma_{-}(-\vec  k_{2} )
\label{p_ph-p_ph}
\end{gather}

 Relation (\ref{10}) is the basic formula of asymmetry in two parity-conjugated processes. The  research team of Harada and his colleagues \cite{Harada} considered the pion-photon scattering with pion as a virtual particle in s-channel (cf. (\ref{11})). They obtain interesting expression of asymmetry. There theta is the angle between pion and photon in the final state. In approximation of large energy, zero limit of the angle, value of energy about 1 GeV and chemical potential about 200 MeV, the maximum value of the asymmetry is equal to zero point two.
\begin{equation}
\mathcal{A}=\left\vert
\cfrac{M^2_{+}-\mathcal{P}[M_{+}]^2}
{\sum_{m=\pm}
\left(
M^2_{m}+\mathcal{P}[M_{m}]^2
\right)}
\right\vert
\label{10}
\end{equation}
\begin{equation}
\mathcal{A}^{\text{s-channel}}\approx
\frac{\mu _{5}N_{c}}{12\pi^{2}f_{\pi}^{2}}
 \cfrac{
E_{\gamma}\left( E^{2}_{\pi}-m^{2}_{\pi} \right)\sin^{2}\theta
}{
m^{2}_{\pi}+E_{\gamma}E_{\pi}
\left(
1-\sqrt{1-\frac{m^{2}_{\pi}}{E^{2}_{\pi}}}\cos\theta
\right)},
\label{11}
\end{equation}
for $\mu_5 < f_{\pi}$.
The choice of the chiral chemical potential as a zero component of an axial vector field and the temporal gauge of the electromagnetic fields provide the physical amplitudes $
M=M_{\text{even-even}}+M_{\text{even-odd}}$.
%

\begin{figure}[!h]
\center{\includegraphics[width=0.5\linewidth]{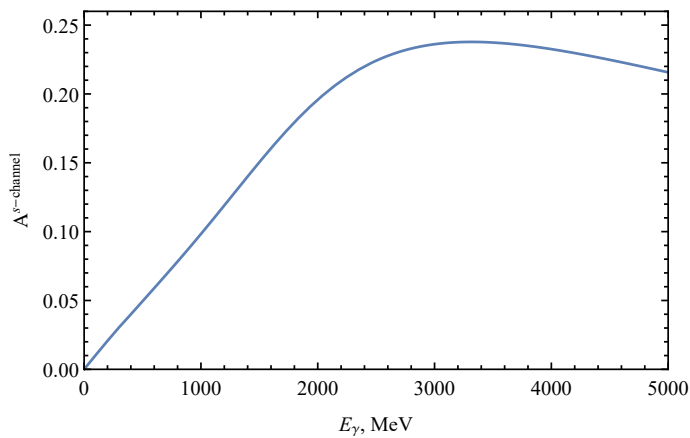}}
\caption{Asymmetry, \(\mu_{5}=100\) MeV, \(E_{\pi2}=1\) GeV}
\label{ris:graph1}
\end{figure}

Fig. \ref{ris:graph1} shows the polarization asymmetry with energy equal to 1 GeV and chemical potential equal 100 MeV. The maximum value of the asymmetry is $\sim 0.2$ .

\section{Conclusions and outlook}

In this work we reported the state-of-art of the study of local parity breaking
(LPB) emerging in a dense hot baryon matter (hadron fireball) in
heavy-ion (central) collisions at high energies. The phenomenology of LPB in a
fireball is based on introducing a topological (axial) charge. Topological charge
fluctuations transmit their influence to hadronic physics via an
axial chemical potential (quark-hadron continuity, see \cite{fuku2018}and refs. therein). We suggested QCD-motivated sigma model for
the description of isotriplet pseudoscalar and isoscalar and
isotriplet scalar mesons  in the medium of a fireball. We conclude:
\begin{itemize}
\item There are two ways to improve the discovery potential: firstly,
to elaborate the recipes for experimentalists to detect  peculiar effects
generated in CP odd background, secondly, to measure the production
of  the mass states without a firm CP parity. In both cases
the chiral chemical potential method helps a lot in predictions.
\item In addition we already suggested \cite{aaep} the vector meson dominance model
with chiral imbalance:
the spectrum of massive vector mesons splits
into three components with different polarizations and with different
effective masses  that can be used to detect local parity breaking.
The proposed schemes for revealing local parity breaking helps to (partially) explain
qualitatively and quantitatively the anomalous yield of
dilepton pairs in the CERES, PHENIX, STAR, NA60, and ALICE
experiments. Accordingly the identification of its physical origin might
serve as a base for a deeper understanding of QCD properties in a
medium under extreme conditions.
\item Some time ago an interesting proposal was set forth in \cite{Harada} to detect the LPB by measuring photon polarization asymmetry in the process $\pi^\pm \gamma \to \pi^\pm \gamma $. We extend this proposal emphasizing the resonance enhancement at  energies comparable with the mass of $\tilde a^\pm_0$ scalars.
\item It is establish that exotic decays of scalar mesons arise as a result of mixing of $\pi$ and $a_0$ vacuum states in the presence of chiral imbalance. The pion electromagnetic formfactor obtains a parity-odd supplement which generates a photon polarization asymmetry in pion polarizability. We consider that the above-mentioned properties of LPB can be discovered in experiments on LHC, RHIC, CBM FAIR and NICA accelerators
\end{itemize}
\section*{Acknowledgements}
It is a pleasure to thank the organizers of the International Conference "Hadron Structure and QCD 2018” dedicated to the memory of Lev N. Lipatov
for a fruitful meeting and an excellent atmosphere. This work has been supported through grants FPA2013-46570, 2014-SGR-104 and Consolider CPAN. Funding was also partially provided by the Spanish MINECO under projectMDM-2014-0369 of ICCUB (Unidad de Excelencia ‘Maria de Maeztu’). A.A. and V.A. were supported by Grant RFBR project 16-02-00348 and also got a financial support of SPbSU, in
particular, by Grants id=27800514; 27803510; 27944831; 28175401; 17652938.

\end{document}